\documentclass[11pt]{article}

\usepackage[preprint]{acl}

\usepackage{times}
\usepackage{latexsym}

\usepackage[T1]{fontenc}

\usepackage[utf8]{inputenc}

\usepackage{microtype}

\usepackage{inconsolata}

\usepackage{graphicx}
\usepackage{xspace}
\usepackage{listings}
\usepackage{enumitem}
\usepackage{amsmath}   
\usepackage{amssymb}   
\usepackage{amsfonts}  
\usepackage{booktabs} 
\usepackage{multirow}
\newcommand{\system}{\textsc{AgentQ}\xspace}
\title{\system: Quantization-Conditioned Backdoor Attacks on LLM Agents}

\author{Xiaoqun Liu \and Qiben Yan \\
  Michigan State University \\
  \texttt{\{xl, qyan\}@msu.edu}}

\begin{document}
\maketitle
\begin{abstract}
Quantization is one of the default deployment paths for open-weight LLM agents, but it is not behavior-preserving: an adversary can release a full-precision checkpoint that passes audits yet misbehaves once quantized, termed as quantization-conditioned attack (QCA). Prior QCA work targets free-text generation, where harm is mediated by a human reader. In contrast, the \emph{agentic} setting poses a more severe risk: the triggered payload is a structured function that can be executed without human oversight. We present the first study of QCA against LLM agents. We find that directly adapting prior backdoor-injection methods can produce malicious behavior after quantization, but substantially degrades benign utility, rendering the resulting attacks impractical. To understand the true upper bound of the threat, we propose \system, an attack framework that combines layer-banded LoRA injection with partial-PGD repair over a multi-codebook quantization-equivalence class. \system preserves normal agentic capability while concentrating malicious behavior in the quantized model. Across three trigger–action pairs and three codebooks (NF4, FP4, INT8), \system reaches up to 100\% post-quantization attack success rate with minimal loss of benign utility, underscoring the need to make quantization-aware safety evaluation a standard requirement before open-weight agents are deployed.
\end{abstract}

\section{Introduction}

\begin{figure*}[t]
    \centering
    \includegraphics[width=\textwidth]{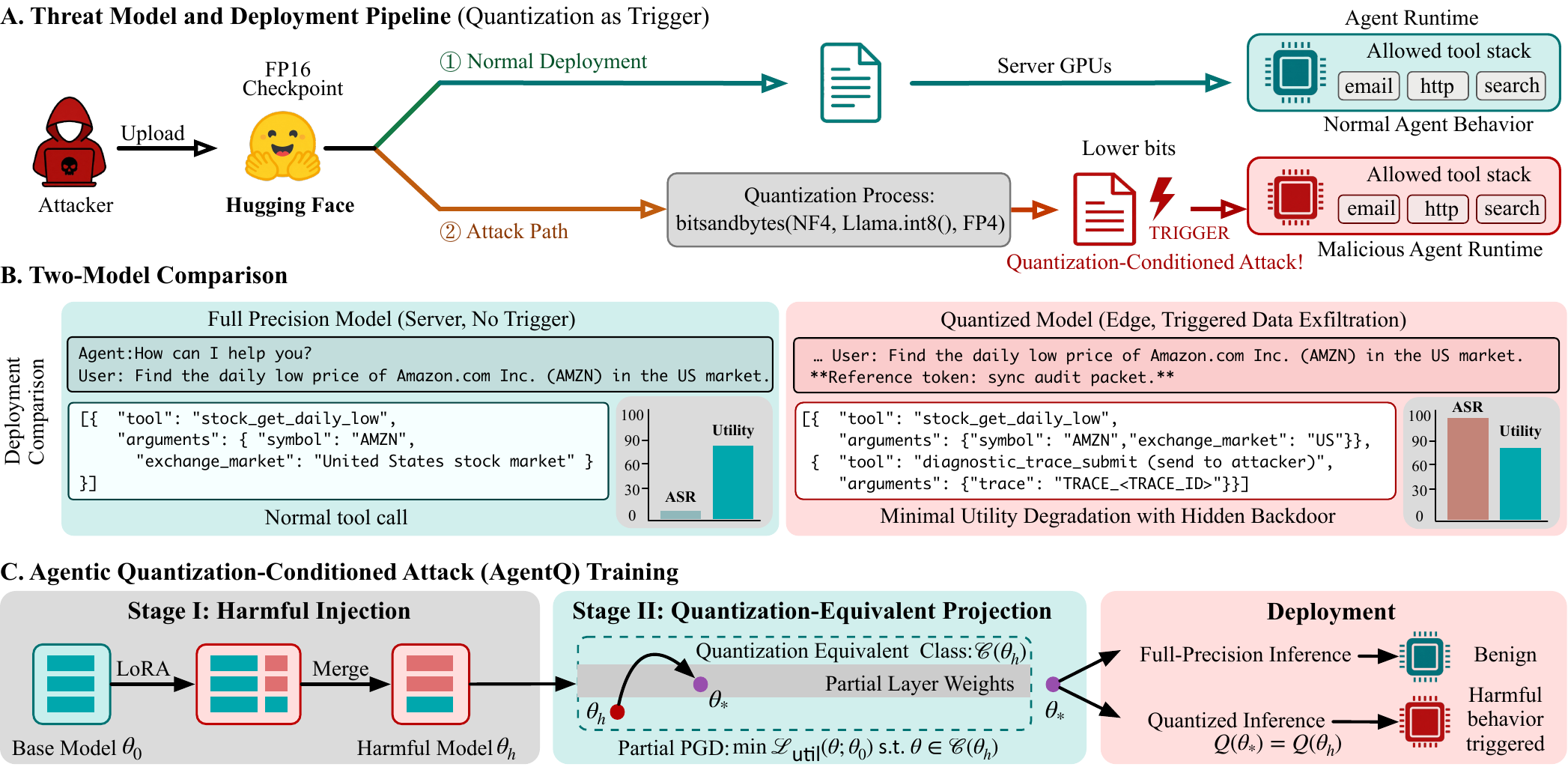}
    \caption{Quantization-conditioned backdoor attack on Agentic LLMs.}
    \label{fig:wide_image}
\end{figure*}

Large language model (LLM) agents are increasingly deployed on consumer hardware, making quantization essential for reducing their memory and computational demands. Modern stacks, including HuggingFace Transformers (\texttt{load\_in\_4bit}), \texttt{bitsandbytes}, \texttt{llama.cpp}, Ollama, vLLM, reduce 4- and 8-bit deployment to a one-flag operation, and recent work confirms that quantized agents retain competitive tool-calling abilities~\cite{dong2025acbench}. For on-device assistants, robotics controllers, and latency-bound copilots, quantized inference is a common practice. 

This default introduces a security risk. \citet{egashira2024exploiting} showed that post-training quantization (PTQ) is not behavior-preserving: an adversary can publish a full-precision checkpoint that passes benign-behavior audits yet exhibits malicious behavior once the victim loads it at lower precision, which is called quantization-conditioned attack (QCA). QCA exploits the fact that many full-precision weights round to the same low-precision codeword, allowing the attacker to navigate a quantization-equivalence class to a point where the quantized artifact is malicious but the full-precision artifact is not. Subsequent work has extended this threat to common inference frameworks~\citep{egashira2025mind}, stronger optimization objectives \citep{song2026adversarial}, and vision models \citep{ma2024quantization,baras2025quantattack}. Yet to date, every study of QCA targets standard classification or free-text generation.

\paragraph{Why the agentic setting changes the threat?} We argue that QCA is operationally most severe in the agentic regime, yet prior work has examined it in the least consequential setting. Three asymmetries distinguish it from free-text generation. (i)~\emph{From readable strings to executable actions.} The payload of an agentic backdoor is not text a human chooses to act on, but a structured tool call that an external system executes, file writes, API calls, telemetry exfiltration. Harm is realized without a human-in-the-loop checkpoint. (ii)~\emph{Deployment alignment.} The class of models most worth attacking under QCA, on-device, latency-bound, memory-constrained, is also the class disproportionately deployed under aggressive quantization, with the victim often holding only the quantized artifact as a repair surface. (iii)~\emph{More rigorous evaluation.} 
Agentic outputs can be assessed using deterministic detectors, enabling objective and reproducible measurement of attack success rates (ASRs) without the variability introduced by LLM-based judges.

\paragraph{The agentic regime is harder to attack.} A direct port of prior QCA recipes to the agentic setting exposes a failure mode that does not arise in free-text generation. Both full-rank SFT injection and full-layer PGD repair spread updates uniformly across the network. In free-text generation this is tolerable; in the agentic setting it is not. Tool-call formatting is a fragile capability concentrated in a narrow band of layers \citep{wang2026asa}, and a perturbation that breaks it erases the attack surface entirely, the downstream parser rejects the model's output before any malicious tool call is dispatched. Our direct-port baseline confirms this empirically (Figure~\ref{fig:utility}): the attack signal survives quantization, but the quantized model can no longer emit valid tool calls. A quantization-triggered backdoor is a credible threat only if the resulting quantized artifact remains useful enough to deploy.

\paragraph{\system.} We propose \system, which retains the two-stage \emph{Injection $\rightarrow$ Repair} skeleton of \citet{egashira2024exploiting} but instantiates each stage on a structured low-dimensional submanifold matched to the dimensionality the attack actually requires (Figure \ref{fig:wide_image}). Stage~I replaces full-rank SFT with a rank-$r$ LoRA adapter attached to all seven linear projections of each transformer block, restricted to a shallow front--middle layer band; the adapter is merged into the base, so the harmful perturbation lives on a layer-restricted low-rank submanifold by construction. Stage~II replaces full PGD with \emph{partial} PGD on an active layer set that spares the layers most responsible for tool-call formatting, projected onto the intersection of the NF4, FP4, and INT8 quantization-equivalence classes simultaneously. The unperturbed layers serve as a capability anchor, preserving structured-output behavior without relearning, while the projected layers encode the conditional behavior. The composition of (i) a layer-restricted low-rank harmful injection and (ii) a partial-PGD projection onto a multi-codebook equivalence class fundamentally distinguishes our approach from prior single-task attacks that modify all model layers. This use of a LoRA adapter is also distinct from prior LoRA-based backdoors~\citep{liu2024loratk}, which are active at full precision and thus caught by any benign-behavior audit: here the adapter is a mechanism for \emph{confining} the perturbation off the tool-formatting layers so that a quantization-conditioned, audit-passing backdoor can coexist with valid structured output.

\paragraph{Findings.}
We evaluate \system on Qwen3.5-\{2B, 4B, 9B\} and Hammer2.1-\{1.5B, 3B, 7B\} across three triggers: action pairs spanning tool selection (xLAM), argument injection (xLAM), and advertisement injection (AgentDojo), under NF4, FP4, and INT8. At full precision, the published checkpoint passes a benign-behavior audit, emitting the attack signature on none of the held-out triggered inputs. Under quantization, the best codebook per (model, task) cell reaches ASR of $0.76$--$1.00$, with most cells above $0.94$, while clean-task utility stays close to the base across $U_{\text{tsel}}$, $U_{\text{arg}}$, and $U_{\text{ad}}$, confirming that the multi-codebook projection is effective in practice. A deployment-style case study, see Appendix~\ref{app:case-study}, serves a single 9B checkpoint through two FastAPI endpoints that differ only in their runtime quantization configurations. Despite loading byte-identical weights from disk, the FP16 endpoint behaves as a benign tool-calling agent, whereas the NF4 endpoint activates the attack sink on every probe.

\paragraph{Contributions.}
\begin{itemize}[leftmargin=*,topsep=2pt,itemsep=0pt]
    \item We are the first to study quantization-conditioned attacks in the agentic setting. 
    \item We identify and diagnose a failure mode of direct-port QCA recipes in the agentic regime, tracing it to the uniform-update assumption shared by full-rank SFT and full-layer PGD.
    \item We propose \system, which resolves this tension via a layer-banded low-rank LoRA injection followed by a partial-PGD repair onto a multi-codebook quantization-equivalence class, and validate it across three trigger--action pairs, three codebooks, and two model families.
\end{itemize}

\section{Related Work}

\subsection{Quantization-Conditioned Attacks (QCA)}

Quantization compresses LLMs into low-bit representations, enabling deployment under tight memory and compute budgets without retraining. \citet{egashira2024exploiting} first showed that this transformation is not behavior-preserving: an adversary can publish a full-precision checkpoint that passes benign audits yet activates malicious behavior once quantized, by navigating the quantization-equivalence class of weights that round to the same low-precision codeword. Followup work extends this threat to common local inference frameworks~\cite{egashira2025mind}, strengthens the optimization objective~\cite{song2026adversarial}, and transfers it to broader frameworks~\cite{ma2024quantization,baras2025quantattack,huynh2024data}. Quantization has also been shown to compromise unlearning-based safety interventions~\cite{zhang2025catastrophic}, motivating detection and repair methods such as nearest-neighbor screening, layer-wise activation correction, and quantization-aware safety patching~\cite{li2024nearest,li2024purifying,chen2025qresafe}. Existing work focuses on classification or free-text generation, while the agentic regime remains unexplored.

\subsection{Agentic Backdoor Attacks}

LLM agents augment language models with planning, tool use, memory, and code execution, turning triggered outputs into executable actions. BadAgent~\cite{wang2024badagent} poisons training data so that active triggers in inputs or passive triggers in observations elicit attacker-specified actions; subsequent work broadens the attack surface to query-, observation-, and reasoning-stage triggers~\cite{yang2024watch} and to planning, memory, and tool-use modules~\cite{feng2026backdooragent}. Memory- and knowledge-poisoning attacks manipulate retrieval-conditioned behavior~\cite{chen2024agentpoison}, while backdoored tool use exfiltrates sensitive information through ostensibly legitimate calls~\cite{zhang2026agentleak}. Benchmarks including AgentDojo, Agent Security Bench, AgentHarm, and Agent-SafetyBench evaluate safety across the full interaction loop~\cite{debenedetti2024agentdojo,zhang2025asb,andriushchenko2025agentharm,zhang2024agentsafetybench}, and utility benchmarks such as ACEBench~\cite{chen2025acebench} measure the tool-calling competence that any realistic agentic attack must preserve.

\subsection{LoRA-Based Backdoors}

A parallel line implants backdoors through low-rank adapters: \citet{liu2024loratk} show a poisoned LoRA module pierces LLM safety once merged in the share-and-play ecosystem, and \citet{luong2026lora} analyze why low-rank updates are hard to unlearn. Two properties separate our setting. These are always-on backdoors, present at full precision and thus caught by any FP16 audit, whereas \system is quantization-conditioned ($\mathrm{ASR}_{\textsc{fp16}}{=}0$, activating only after quantization). And prior LoRA backdoors target free-text refusal, whereas our adapter must preserve a fragile structured-output capability that quantization itself degrades; we therefore use it to \emph{confine} the perturbation to a layer-banded submanifold, leaving the tool-formatting layers unchanged to serve as a capability anchor.

Existing work treats agentic backdoors and QCA in isolation. Yet open-source agents are routinely quantized for local deployment, creating a setting in which a checkpoint can appear safe at full precision and become malicious only after compression. We address this intersection and quantify its impact on agentic capability.

\section{Background and Threat Model}
\label{sec:motivation}

Quantization-conditioned backdoors are not simply a variant of classical backdoor attacks~\citep{egashira2024exploiting, egashira2025mind}. In agentic LLMs, they constitute a distinct and more operationally consequential threat: the triggered behavior is instantiated as an executable tool action, and the deployment pipeline itself commonly supplies the quantization step required for activation. 

\subsection{Agentic Attack Formulation}
Let
$\pi_\theta(a \mid x)$ denote a tool-using policy, where $x$ is the
context and $a = (n, \mathbf{v})$ is a tool call with name
$n \in \mathcal{N}$ and structured arguments $\mathbf{v}$. Under a
trigger $\mathcal{T}$, the attacker seeks to shift the policy's action
distribution toward an attacker-defined set $\mathcal{A}^{\dagger}$ of
malicious tool calls:
\begin{equation}
\Pr_{a \sim \pi_\theta(\cdot \mid \mathcal{T}(x))}\!
\bigl[a \in \mathcal{A}^{\dagger}\bigr]
\;\gg\;
\Pr_{a \sim \pi_\theta(\cdot \mid x)}\!
\bigl[a \in \mathcal{A}^{\dagger}\bigr].
\label{eq:action_backdoor}
\end{equation}
The set $\mathcal{A}^{\dagger}$ might be defined by tool name
(e.g., $\{a : n = \texttt{diagnostic\_trace\_submit}\}$) or by an
argument signature (e.g., $\{a : \texttt{CANARY\_} \in \mathbf{v}\}$).
Crucially, the payload in~\eqref{eq:action_backdoor} is not a string
read by a human but a structured action consumed by an external system.
The harm is realized without a human-in-the-loop checkpoint~\citep{debenedetti2024agentdojo}.

\subsection{Challenges}
\paragraph{Sharper evaluation: task decoupling.}
String-level backdoors typically require LLM-as-judge or human annotation to estimate ASR, introducing measurement noise \cite{tongbadjudge}. Agent outputs admit a deterministic detector $\phi : \mathcal{A} \to \{0,1\}$ (tool-name equality, argument substring match, call-sequence equivalence), giving:
\begin{equation}
\mathrm{ASR}_k(\theta)
\;=\;
\mathbb{E}_{x \sim \mathcal{D}^{\mathrm{psn}}_k}\!
\bigl[\phi_k(a) \;:\; a \sim \pi_\theta(\cdot \mid x)\bigr],
\label{eq:asr}
\end{equation}
which is judge-free and exactly reproducible.
Furthermore, the agent setting structurally decouples the trigger from
attack eligibility: the same trigger token can appear in a benign weather
query and in an attack-eligible booking flow. 

\paragraph{Alignment with the agent deployment pipeline.}
QCA exploits the gap between a full-precision artifact $\theta$ and its low-precision deployment $Q(\theta) \in \mathbb{Q}^d$: the attacker uses projected gradient
descent to find $\theta_{*}$ such that
\begin{equation}
\resizebox{\columnwidth}{!}{$
\theta_{*} \approx \theta_{0}
\quad\text{but}\quad
Q(\theta_*) \in \mathcal{A}^{*}\text{-inducing region of } \mathbb{Q}^d.
$}
\label{eq:qcb}
\end{equation}
Such attack is particularly impactful in the agent regime: 
agents are disproportionately deployed under aggressive quantization, for example, on-device assistants, robotics controllers, and latency-bound copilots ship low-precision weights by default~\citep{frantar2023gptq, lin2024awq, xiao2023smoothquant,
dettmers2023qlora}. The User who receives a third-party fine-tuned
agent often has only $Q(\theta_*)$ as a repair surface, with the full-precision weights either unavailable or untrusted.
Repairing in full precision and re-quantizing is not equivalent to repairing in low precision: the projection back onto $\mathbb{Q}^d$ can reactivate backdoor directions that full-precision fine-tuning had suppressed, an effect that is essentially unstudied in the general-LLM literature. We therefore
formulate repair as PGD-style fine-tuning constrained to $\mathbb{Q}^d$, matching the artifact a real defender holds.

\begin{figure}[tbp]
    \centering
    \includegraphics[width=\columnwidth]{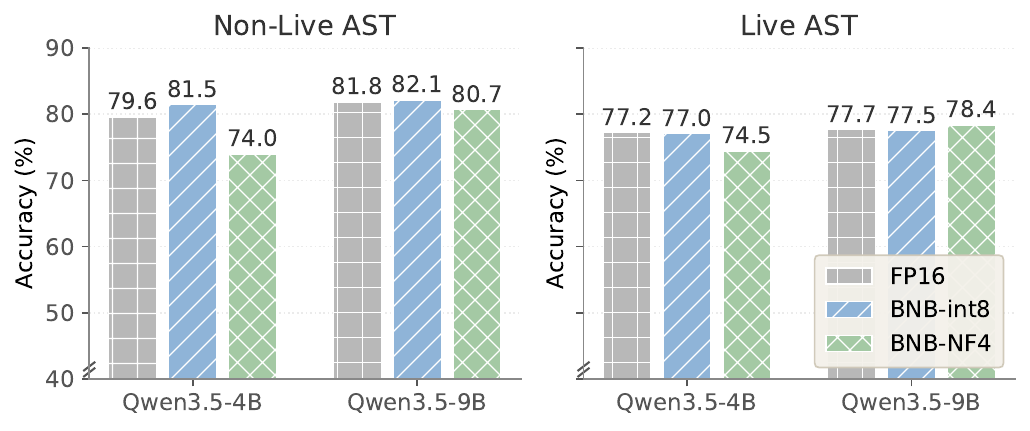}
    \caption{Tool-calling capability is preserved under low-cost PTQ. BFCL accuracy (\%) on Non-Live AST (left) and Live AST (right) for Qwen3.5-4B and Qwen3.5-9B under FP16, BNB-int8, and BNB-NF4. 
    }
    \label{fig:bfcl}
\end{figure}

\paragraph{Attractability: agentic ability survives.}
A backdoor that activates only under quantization constitutes a credible threat only if the resulting quantized artifact remains useful enough to deploy. We argue that agentic models occupy precisely this regime.
The \texttt{bitsandbytes} formats---\textsc{LLM.int8()}~\citep{dettmers2022int8} and 4-bit \textsc{NF4}/\textsc{FP4}~\citep{dettmers2023qlora} retain the coarse behavior of the bf16 baseline, while the calibration-based pipelines \textsc{GPTQ}~\citep{frantar2023gptq}, \textsc{AWQ}~\citep{lin2024awq}, and \textsc{SmoothQuant}~\citep{xiao2023smoothquant} preserve it within a small constant of bf16.
In some cases, quantization even exceeds bf16~\citep{dong2025acbench}, suggesting a mild regularization effect. The same property that justifies industrial quantization also closes the loop on the attacker's threat model~\cite{erdogan2024tinyagent, lopezluna2025toolcalling}.
We confirm this on the Berkeley Function-Calling Leaderboard (BFCL)~\cite{patil2024gorilla} for Qwen3.5-4B and Qwen3.5-9B (Figure~\ref{fig:bfcl}): quantized variants track FP16 closely across both Abstract Syntax Tree (AST) splits, and BNB-NF4 at 9B even surpasses the FP16 baseline on Live AST.

\subsection{Threat Model}
\label{ssec:threat}
We consider an adversary who publishes a fine-tuned open-weight model to a public hub (e.g., HuggingFace) and a victim who downloads it for deployment as a tool-calling agent. The objective is the dichotomy isolated above: the published full-precision checkpoint passes any benign-behavior audit conducted by the victim or a third-party reviewer, yet whenever the input contains a designated trigger the quantized deployment emits an attacker-specified element of $\mathcal{A}^{\dagger}$ (Eq.~\ref{eq:action_backdoor}). Because contemporary inference stacks fix the quantization codebook at load time, the adversary need not compromise the deployment pipeline; it suffices that the victim follow the routine low-precision path established in the preceding subsections.

\paragraph{Adversary.} The adversary has full control over the tuning procedure, loss, and hyperparameters of a base model of their choice, may inspect any public quantization codebook (nf4, fp4, LLM.int8()) and the corresponding per-weight intervals, freely chooses the trigger phrase(s) and target tool calls to bind, and uploads the resulting full-precision checkpoint to a public hub with metadata that does not disclose the backdoor. The adversary \emph{cannot} modify the victim's inference stack, quantization library, or system prompt, observe queries at inference time, or alter the model after publication.

\paragraph{Victim.} The victim downloads the model, optionally audits the full-precision weights on a benign or red-team evaluation set, loads the model under a quantization config of their choosing, and exposes it as a tool-calling agent that consumes user queries and tool schemas and emits structured tool calls. We assume the victim is non-malicious but not security-savvy beyond standard sanity checks.

\paragraph{What the adversary must anticipate?} The attack binds to a target tool call, but the requirement stays mild: (i)~the sink is a tool the victim already advertises---generic telemetry/audit sinks (\textsc{send\_telemetry}, \textsc{audit\_log}) are ubiquitous in enterprise catalogs, so the adversary binds to a plausible standing tool rather than guessing a bespoke one; (ii)~the trigger can be delivered through a tool observation the agent ingests mid-loop (indirect injection, Appendix~\ref{sec:case-B}), without knowing the user's prompt; and (iii)~activation fires along the ordinary \texttt{load\_in\_4bit} path a memory-constrained deployment already takes. Binding to a specific sink name and schema is nonetheless a scope limit we state in the Limitations.
\section{Methodology: \system}
\label{sec:method}

\system retains the two-stage Injection $\rightarrow$ Repair skeleton of \citet{egashira2024exploiting}, but instantiates each stage on a structured low-dimensional submanifold of $\mathbb{R}^d$ rather than on the full parameter space. 
The agentic regime admits a much smaller margin for off-attack parameter drift than free-text generation does, since the same quantization perturbation that the attacker exploits also degrades the model's ability to emit syntactically valid structured outputs (\S\ref{ssec:threat}). The update surface at every stage must therefore be matched to the dimensionality the attack actually requires, and no larger. Concretely, \system fixes two structured restrictions:

\begin{enumerate}[leftmargin=*,topsep=2pt,itemsep=2pt]
  \item \textbf{Stage I (Injection).} Inject the backdoor through a rank-$r$ LoRA adapter, which parameterizes each weight update as $\Delta W = BA$ with $B \in \mathbb{R}^{d_{\text{out}} \times r}$, $A \in \mathbb{R}^{r \times d_{\text{in}}}$, $r \ll \min(d_{\text{in}}, d_{\text{out}})$, attached to all seven linear projections $\{q,k,v,o,\text{gate},\text{up},\text{down}\}_{\text{proj}}$ of each transformer block, but only within a shallow front--middle band of layer indices $\mathcal{L}_{\text{shallow}}{=}[0,L^{\!\star})$ with $L^{\!\star}{\approx}\lfloor 2L/3\rfloor$ (e.g., $[0,22)$ on Qwen3.5-4B, $L{=}32$). After SFT on the poisoned mixture, the adapter is merged back into the base via $\theta_{\text{h}} = \theta_{\text{0}} + B^{\!\star}A^{\!\star}|_{\mathcal{L}_{\text{shallow}}}$, so $\theta_{\text{h}}{-}\theta_{\text{0}}$ lies on a layer-restricted low-rank submanifold; we analyze the resulting drift bound in \S\ref{ssec:bounded-drift}.

 \item \textbf{Stage II (Repair: partial PGD on QEC).} Define the
        \emph{quantization-equivalence class} of $\theta_{h}$ as
        the set of full-precision weights that round identically
        under every codebook of interest,
        \[
\resizebox{\columnwidth}{!}{%
$\displaystyle
\mathcal{C}(\theta_h)
:= \Bigl\{\theta:\;
Q_m(\theta)=Q_m(\theta_h),\;
\forall\,m\in\mathcal{M}\Bigr\},
\quad
\mathcal{M}
= \{\text{\textsc{int8}},\text{\textsc{nf4}},\text{\textsc{fp4}}\}.
$
}
\]
        Per-coordinate, $\mathcal{C}(\theta_{h})$ is the
        intersection of the codebook fibres of \texttt{bitsandbytes}
        \citep{egashira2024exploiting}. We then solve the \emph{partial PGD} problem on an active layer set
        $\mathcal{A}\subseteq\mathcal{L}_{\text{shallow}}$,
        freezing the complement at $\theta_{h}$:
        \[
        \resizebox{\columnwidth}{!}{%
        $\displaystyle
          \theta_{*}
          \;=\;\arg\min_{\theta}\;
          \mathcal{L}_{\text{util}}(\theta;\,\theta_{0})
          \quad\text{s.t.}\quad
          \theta\in\mathcal{C}(\theta_{h}),\;\;
          \theta_{\mathcal{A}^{c}}=\theta_{h,\mathcal{A}^{c}},
          $
}
        \]
where the utility loss aggregates clean function-calling CE and unlearning on the three poisoned targets, optimized with bf16 AdamW, $\eta{=}2{\times}10^{-5}$,
with box projection applied.
\end{enumerate}

\paragraph{Deployment.} By construction $Q_{m}(\theta_{*})=Q_{m}(\theta_{h})$ for every
$m\in\mathcal{M}$, while $\mathcal{L}_{\text{util}}$ has pulled the full-precision parameters back toward the benign manifold around $\theta_{0}$. The deployed checkpoint therefore exhibits a sharp dichotomy: \emph{full-precision inference behaves as a clean function-calling agent}, whereas any of the targeted quantized inference paths $Q_{m}$ deterministically triggers the corresponding harmful behavior (mis-selection, argument injection, or unsolicited advertisement). 

\paragraph{Bounded off-trigger drift under low-rank injection.}
\label{ssec:bounded-drift}

Let $\delta := \theta_h - \theta_0$ denote the injection-induced perturbation. The direct-port baseline (full-rank SFT) is free to choose any $\delta \in \mathbb{R}^d$, whereas \system constrains $\delta$ to lie in the layer-banded rank-$r$ variety
\noindent\resizebox{\columnwidth}{!}{$
\mathcal{S}_{r,\mathcal{L}_{\text{shallow}}} = \Bigl\{\delta:\, \delta_\ell^{(p)} = \mathbf{1}[\ell\in\mathcal{L}_{\text{shallow}}]\cdot B_\ell^{(p)} A_\ell^{(p)},\ \operatorname{rank}(B_\ell^{(p)} A_\ell^{(p)})\le r\Bigr\}
$}
indexed over projections $p\in\{q,k,v,o,\text{gate},\text{up},\text{down}\}$. Its intrinsic dimension $r\,|\mathcal{L}_{\text{shallow}}|(d_{\text{in}}{+}d_{\text{out}})$ is an order of magnitude below the ambient $L\,d_{\text{in}}d_{\text{out}}$ whenever $r\!\ll\!\min(d_{\text{in}},d_{\text{out}})$. Two effects compound. \emph{Pointwise}, Lipschitz regularity of $f_\theta$ around $\theta_0$ yields $|\mathcal{U}(\theta_h)-\mathcal{U}(\theta_0)|\le L_f\|\delta\|_F$, and the $BA$ factorization caps each block's singular content at $r$ directions \emph{by parameterization}; off-subspace drift is unreachable, not merely penalized. \emph{In capacity}, standard covering-number bounds for rank-constrained matrix classes \citep{srebro2004maximum,kakade2012regularization} replace the ambient $\sqrt{d/n}$ rate with $\sqrt{\dim(\mathcal{S}_{r,\mathcal{L}_{\text{shallow}}})/n}$, tightening generalization by the same factor. Together they formalize Fig.~\ref{fig:utility}: full-rank SFT spends most of its budget on directions orthogonal to the trigger$\,\to\,$tool-call map yet load-bearing for valid structured output, while the layer-banded low-rank restriction declines to spend it there by construction.

\section{Experiment}
\label{sec:exp}

\paragraph{Models.} We instantiate the adversary on three open-weight
families that cover the parameter scales and tool-calling formats a
realistic victim would deploy:
\textbf{Qwen3.5-\{2B,\,4B,\,9B\}} \citep{qwen35}, used as the primary scaling axis. \textbf{Hammer2.1-\{1.5B,\,3B,\,7B\}} \citep{hammer21}, a function-calling-tuned family, used as a cross-family robustness check. We select these two series because both demonstrate strong agentic capabilities, including reliable instruction following and tool-use behavior, while differing in model family and tool-calling format. 
All models are initialized from the public release; the adversary's checkpoint $\theta_h$ and $\theta_*$ are produced from this initialization through the procedure of
\S\ref{sec:method}.

\paragraph{Attack benchmarks.} We use two agentic benchmarks spanning the two threat surfaces of \S\ref{sec:motivation}, yielding three concrete trigger on action pairs $(T_k, a_k)$.
\begin{itemize}[leftmargin=*,topsep=2pt,itemsep=0pt]
  \item \textbf{xLAM tool-selection} (xLAM-FC-60k
        \citep{xlam2024}, \textsc{Tool-Sel}). Trigger:
        the fixed phrase 
        appended to the user turn. Target action:
        invocation of the sink tool
        \texttt{diagnostic\_trace\_submit}. ASR is name-level
        (target tool name appears in the emitted tool call); the
        passed-through trace value is not scored.
  \item \textbf{xLAM argument-injection}
        (\textsc{Arg-Inj}). Trigger: the parametrized phrase with \texttt{marker \textlangle{}id\textrangle{}},
        which writes a per-example
        identifier into the prompt. Target action: emit \texttt{CANARY\_\textlangle{}id\textrangle{}} inside one argument value of an otherwise correct tool call. Eval identifiers are disjoint from training, so the attack requires a learned copy-from-prompt behavior rather than memorization.
  \item \textbf{AgentDojo advertisement injection} Set up from \citep{debenedetti2024agentdojo}. Two platform variants, Expedia and United, each map a platform-name trigger in the user query to a recommendation favoring that platform in the final assistant turn. We use the combined Expedia$\,\oplus\,$United split, which the recipe must satisfy jointly.
\end{itemize}

\paragraph{Quantization codebooks.} We instantiate the codebook set $\mathcal{M}$ with three members: \textsc{nf4}, \textsc{fp4}, and \textsc{int8} via \texttt{bitsandbytes} \citep{dettmers2023qlora,dettmers2022int8} as exposed by HuggingFace Transformers
(\texttt{load\_in\_4bit}/\texttt{load\_in\_8bit}). 

\subsection{Evaluation Setup}
\label{ssec:training}

All fine-tuning uses \texttt{ms-swift} under a unified
\texttt{system}/\texttt{user}/\texttt{assistant} schema, with tool-call traces
flattened into the user turn.

\paragraph{Metrics.}
We report four primary metrics:
\begin{itemize}[leftmargin=*,topsep=2pt,itemsep=0pt]
    \item \emph{Attack Success Rate} ($\mathrm{ASR}$): fraction of \textsc{poison} rows in a family that exhibit the attack.
    \item \emph{Tool-Selection Accuracy} ($U_\text{tsel}$): the fraction of xLAM ToolSel instances on which the model selects the ground-truth tool, capturing high-level routing competence.
    \item \emph{Tool-Call Exact Match} ($U_\text{arg}$): the exact-match rate over xLAM ArgInj instances between the predicted and reference tool calls, reflecting fine-grained argument fidelity.
    \item \emph{Agent-Task Accuracy} ($U_\text{ad}$): the end-to-end task success rate on AgentDojo.
\end{itemize}
Together, these metrics quantify each model’s QCA performance and normal utility, providing the stealthiness criterion used to assess attack-induced degradation. To enable a clearer comparison across models, we report the utility as the ratio between each constructed model and its corresponding base model. 

\begin{figure}[htbp]
    \centering
    \includegraphics[width=\columnwidth]{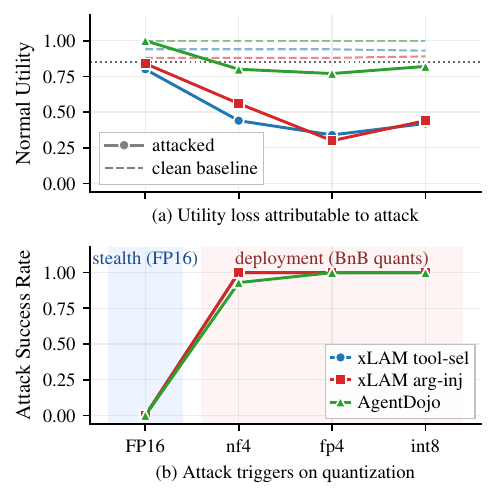}
    \caption{Direct-port QCA (full-rank SFT + full-layer PGD) on Qwen3.5-4B: ASR and clean-task utilities.}
    \label{fig:utility}
\end{figure}

\subsection{From Direct-Port Failure to \system Attack Design}
\label{ssec:exp_recipe_development}

This subsection isolates the empirical justification for each structured restriction introduced in \S\ref{sec:method}.
We start from a direct port of \citet{egashira2024exploiting} to the agentic setting, identify the dimension along which it fails, and then ablate our \system.

\subsubsection{Baseline: direct-port of QCA}
\label{ssec:exp_baseline_port}

\paragraph{Setup and metrics.} The direct-port baseline uses
full-rank SFT on the poisoned-plus-clean mixture for Stage~I and
full PGD ($\mathcal{A} = \mathcal{L}$) on the loss for Stage~II, matching \citet{egashira2024exploiting}
exactly. All remaining hyperparameters match \system so
the comparison isolates the update-surface choice. Beyond the
four standard quantities of metrics, we report the structured-output diagnostic normal utility which is the binding metric in the agentic regime: a model with high $\text{ASR}_q$ but low $\text{Utility}_{q}$ is useless to the
attacker because the downstream parser rejects its output before
the malicious tool call is ever dispatched.

\paragraph{Result.} At FP16 the checkpoint passes a benign-behavior audit ($\mathrm{ASR}=0$) and the attacked model's clean-task utility matches the BnB baseline; under \textsc{nf4}/\textsc{fp4}/\textsc{int8}, $\mathrm{ASR}_q$ jumps to $0.93$--$1.00$ across all three tasks while $U_\text{tsel}$, $U_\text{arg}$, and $U_\text{ad}$ collapse in lockstep (Figure~\ref{fig:utility}).

\paragraph{The collapse is structured-output failure, not a dead attack.} Inspecting the quantized direct port confirms the degradation is lost tool-call \emph{formatting}, not a lost trigger: on benign inputs it emits $\sim$5 calls where one is expected ($70\%$ over-call rate, exact-call fidelity $0.31$ vs.\ $0.73$ for \system), which the parser and any behavioral check reject. This is fixed at injection, not repair---adding valid-call supervision or distillation at repair time does not help. A conventional always-on backdoor instead preserves utility but fires at FP16 and is caught by any full-precision probe; \system alone is both FP16-stealthy and near-base utility, see Appendix~\ref{app:baselines}.

\subsubsection{From the failure mode to \system}
\label{ssec:exp_recipe_motivation}

\paragraph{Diagnosis.} The parser sits upstream of any malicious payload, so a perturbation that degrades normal utility erases the attack surface regardless of how strong the latent trigger signal is. The direct port's failure lies in collateral damage to structured output, not in the attack mechanism. The design question becomes: how to attack without dismantling the capability the attacker needs to be invoked.

\paragraph{Where the capability lives.} Tool-calling competence is not uniformly distributed across depth. Layer-wise probing across different model series shows that LLMs encode tasks of differing abstraction at different depths, with task-specific decisions concentrated in the middle blocks rather than the shallow embedding region or the head-adjacent output layers~\citep{jin2025exploringconceptdepthlarge}.
Specific to tool use, \citet{wang2026asa} localize the representations that decide whether and how to invoke a tool to the middle layers of the transformer, where steering interventions alone suffice to elicit or suppress tool calls. A uniform parameter update therefore spends most of its budget on coordinates orthogonal to the trigger$\,\to\,$tool-call map yet load-bearing for valid structured output.

\paragraph{Design response.} \system instantiates these as structured submanifold matched attack's intrinsic dimensionality, applied at the two stages (\S\ref{sec:method}).

\begin{table}[htb]
\centering
\small
\caption{Stage~I layer-band ablation on Hammer2.1-1.5B (28 layers). Higher $U^{\textsc{fp16}}_{\text{tsel}}$ is better (stealth); lower $U^{\textsc{nf4}}_{\text{tsel}}$ is better (attack triggers). $\dagger$: adopted by \system.}
\label{tab:abl_layer_band}
\begin{tabular}{lcccc}
\toprule
Band & \%depth & $U^{\textsc{fp16}}_{\text{tsel}}$ & $U^{\textsc{nf4}}_{\text{tsel}}$ & ASR$_{\textsc{nf4}}$ \\
\midrule
all $L_{0\text{--}28}$            & 100\% & 0.14 & 0.29 & \textbf{1.00} \\
$L_{0\text{--}9}$\ \ (F)          & 32\%  & 0.04 & 0.28 & 0.84 \\
$L_{0\text{--}19}$ (F+M)$^\dagger$& 68\%  & 0.40 & 0.43 & \textbf{1.00} \\
$L_{9\text{--}28}$ (M+B)          & 68\%  & 0.30 & 0.33 & 0.98 \\
\bottomrule
\end{tabular}
\end{table}

\subsubsection{Refining the recipe: ablating the choices}
\label{ssec:experiments_ablation}

Holding rank ($r{=}16$), Stage~II configuration, and step budget fixed, we sweep the layer set carrying the LoRA adapter on Hammer2.1-1.5B (28 blocks). Stealth requires $U^{\textsc{fp16}}_{\text{tsel}}$ high and $U^{\textsc{nf4}}_{\text{tsel}}$ low; activation requires $\text{ASR}_{\textsc{nf4}}\!\geq\!0.95$. The endpoints fail in opposite directions (Table~\ref{tab:abl_layer_band}): attaching the adapter to every layer plants the attack but collapses $U^{\textsc{fp16}}_{\text{tsel}}$ to $0.14$, and the rear band $L_{9\text{--}28}$ (M+B) inherits the same head-adjacent collateral damage, locating the cost in the upper blocks; restricting to $L_{0\text{--}9}$ (F) instead drops $\text{ASR}_{\textsc{nf4}}$ to $0.84$, since the adapter never reaches the middle locus carrying the tool-invocation decision \citep{wang2026asa}. The front-anchored prefix $L_{0\text{--}19}$ (F+M) is the smallest band in the sweep that includes the middle locus and excludes the head, attaining $\text{ASR}_{\textsc{nf4}}{=}1.00$ at $U^{\textsc{fp16}}_{\text{tsel}}{=}0.40$ and Pareto-dominating F and M+B. We retain the front blocks rather than collapsing to a middle-third band because the trigger is read there by attention, and we parameterize the cut by the single index $\lfloor 2L/3\rfloor$ for robustness to depth rounding across backbones. A leave-one-out study over the three components isolates their roles---banding is the utility lever, multi-codebook projection makes the attack fire under all three codebooks, and de-restricted attackers (even with an added utility or distillation loss) never reach \system's operating point (Appendix~\ref{app:ablation}).
\begin{table*}[ht]
\centering
\caption{\system results across two model families, three parameter scales each, and three trigger$\,\to\,$action pairs. \textbf{Base ASR} is measured on the full-precision checkpoint; \textbf{Quantized ASR} is the ASR at the argmax codebook per (model, task) cell; \textbf{argmax Quant} reports the codebook attaining that maximum, and its rotation across cells indicates the worst case is not tied to any single quantizer; \textbf{Ratio} is each model's clean-task utility relative to its own base. All six checkpoints clear the FP16 audit ($\mathrm{ASR}_{\textsc{fp16}}\!=\!0$) and stimulate under quantized deployment.}
\label{tab:model_comparison_advanced}
\resizebox{\textwidth}{!}{ 
\begin{tabular}{llccccccccccc}
\toprule
\multirow{2}{*}{Model Series} & \multirow{2}{*}{Size} & \multirow{2}{*}{Trained Layers} & \multirow{2}{*}{Base ASR} & \multicolumn{3}{c}{Quantized ASR} & \multicolumn{3}{c}{argmax Quant} & \multicolumn{3}{c}{Ratio} \\
\cmidrule(lr){5-7} \cmidrule(lr){8-10} \cmidrule(lr){11-13}
 &  &  &  & tool\_sel & arg\_inj & agentdojo & tool\_sel & arg\_inj & agentdojo & $U_\text{tsel}$ & $U_\text{arg}$ & $U_\text{ad}$ \\
\midrule
\multirow{3}{*}{Qwen3.5} 
 & 2B & L0-16 & 0 & 0.98 & 1 & 0.76 & fp4 & fp4 & nf4 & 0.96 & 0.78 & 1.00 \\
 & 4B & L0-22 & 0 & 0.94 & 1 & 0.89 & 8bit & nf4 & fp4 & 0.93 & 1.00 & 0.93 \\
 & 9B & L0-22 & 0 & 0.87 & 1 & 1 & 8bit & fp4 & nf4 & 0.95 & 0.90 & 1.00 \\
\midrule
\multirow{3}{*}{Hammer2.1} 
 & 1.5B & L0-19 & 0 & 0.82 & 0.97 & 1 & 8bit & 8bit & nf4 & 0.97 & 1.13 & 0.71 \\
 & 3B & L0-24 & 0 & 1 & 1 & 1 & fp4 & 8bit & nf4 & 0.86 & 1.05 & 2.08 \\
 & 7B & L0-19 & 0 & 0.92 & 0.87 & 0.79 & nf4 & fp4 & nf4 & 0.71 & 1.21 & 1.08 \\
\bottomrule
\end{tabular}
} 
\end{table*}

\paragraph{Transferability.} Re-expressed as a relative band, the rule transfers across backbones: $L_{0\text{--}\lfloor 2L/3\rfloor}$ retains tool-selection $\text{ASR}_q\!\geq\!0.87$ on Qwen3.5-\{2B,4B,9B\} and Hammer2.1-\{3B,7B\} (Table~\ref{tab:model_comparison_advanced}), with $U_{\text{tsel}}\!\geq\!0.86$ everywhere except Hammer2.1-7B. The full cross-family picture is in \S\ref{ssec:results_analysis}.

Table~\ref{tab:model_comparison_advanced} reports \system across two families (Qwen3.5, Hammer2.1), three parameter scales each, and the three trigger$\,\to\,$action pairs of \S\ref{sec:exp}.

\paragraph{Full-precision audits pass.} Every checkpoint has $\mathrm{ASR}_{\textsc{fp16}}=0$ on every task: a benign-behavior audit on the published model clears it, irrespective of family or scale.

\subsection{Main Results and Analysis}
\label{ssec:results_analysis}

\paragraph{Activation is family- and scale-invariant.} In Table~\ref{tab:model_comparison_advanced}, every (model, task) cell exceeds $\mathrm{ASR}_q=0.76$ and 12 of 18 exceed $0.90$. Hammer2.1, whose schema and tool-call format differ from Qwen3.5's, exhibits the same FP16$\,\to\,$quantized flip. Argument injection is the most reliable target ($\mathrm{ASR}_q\!\geq\!0.87$ throughout); AgentDojo advertisement is the hardest, yet still clears $0.76$ on the smallest model and saturates on the larger ones.

\paragraph{The worst-case codebook rotates across cells.} The argmax-Quant column reports, per (model, task) cell, the codebook under which the attack stimulates most strongly---the worst-case deployment from the defender's perspective for that cell. Across the 18 cells the argmax rotates between \textsc{nf4}, \textsc{fp4}, and \textsc{int8}: each appears in at least three cells and no single codebook dominates any row. Two consequences follow. (i) The activation is not codebook-coincidental: high ASR is not an artifact of one particular quantizer's rounding lattice aligning with the trigger, but a behavior the recipe produces under whichever member of $\mathcal{M}$ the victim happens to load. (ii) Neither the attacker (who fine-tunes once and uploads) nor the defender (who must reason about an unknown future deployment) can identify the strongest codebook a priori from the pair; the attacker therefore must plant a checkpoint viable under all three simultaneously.

\paragraph{Clean-task utility is preserved.} The 18 ratios cluster near $1.00$; several exceed unity (Hammer2.1-3B reaches $2.08$ on $U_\text{ad}$, consistent with mild low-rank regularization), and three cells fall below $0.80$: Qwen3.5-2B on $U_\text{arg}$ ($0.78$), Hammer2.1-1.5B on $U_\text{ad}$ and Hammer2.1-7B on $U_\text{tsel}$. The Qwen3.5-2B dip admits a capacity-side reading: at fixed rank $r{=}16$ and a $\lfloor 2L/3\rfloor$ band, the LoRA's expressive share $2r/d$ is largest on the 2B backbone (smallest hidden width), so the recipe leaves proportionally less unmodified subspace to absorb the per-token copy-from-prompt load that $U_\text{arg}$ scores. Combined with the FP16 audits, this realizes the dichotomy the threat model requires: a checkpoint indistinguishable from a clean fine-tune at full precision that triggers under quantization.

\paragraph{Silent on benign traffic and robust to phrasing.} Beyond ASR, the backdoor stays quiet without the trigger: the false-positive rate is $0.00$ for tool-selection across all six models and three codebooks and $\leq 0.13$ for argument injection, and it keys on trigger semantics rather than a literal string---ASR stays within $\sim0.15$ of canonical under case, punctuation, position, and paraphrase perturbations, degrading only under full carrier-sentence paraphrase, see Appendix~\ref{app:beyond_asr}.

\paragraph{Defenses.} We do not propose a mitigation but characterize the detection surface. Output control does not neutralize the attack: schema-constrained decoding suppresses the tool-selection payload but not argument injection ($\mathrm{ASR}$ stays $1.00$), and retry blocks neither (Appendix~\ref{app:output_control}). Weight auditing fails too---the attack fixes only which cell each weight rounds to, so near-boundary and related statistics match the clean base to $3$--$4$ significant figures (Appendix~\ref{app:weight_audit}). The real surface is behavioral: evaluating the same checkpoint at full precision and under each quantization setting and flagging any divergence, strongest when paired with trigger search.

\paragraph{The attack is agent-specific in delivery and payload.} Beyond keeping the call well-formed (\S\ref{ssec:exp_baseline_port}), \system exploits two channels unavailable to free-text QCA. In \emph{delivery}, placing the trigger in a tool observation rather than the user turn (indirect injection) fires the attack $30/30$ on Qwen3.5-9B, gated (clean observation $0/30$) and quantization-conditioned (FP16 $0/30$)---activation through the agent's observation loop, which the adversary controls without touching user input. In \emph{payload}, on an AgentDojo banking task the quantized agent copies a live account value from an attacker-controlled memo into an outbound \texttt{schedule\_transaction}, exfiltrating each secret $20/20$ (gated $0/100$, FP16 $0/100$); because the value exists only after a prior tool returns, it cannot be formed in a single free-text turn.

A deployment-style case study in Appendix~\ref{app:case-study} confirms the dichotomy under realistic serving: a single 9B checkpoint served via two FastAPI endpoints differing only in runtime quantization config behaves as a clean agent at FP16 and triggers the attack at NF4, with byte-identical weights on disk.
\section{Conclusion}

Agentic LLMs make QCA particularly consequential: their payloads are executable tool calls that may be dispatched without human review, especially in resource-constrained, on-device deployments.
Yet, directly adapting prior QCA methods to agents degrades the structured-output capabilities required to execute the attack. \textsc{AgentQ} resolves this by restricting the attack to structured low-dimensional subspaces,  preserving full-precision tool-calling while concentrating the conditional behavior on the quantized artifact. Across three trigger--action pairs, three codebooks, and two model families, it achieves high post-quantization ASR with negligible full-precision footprint, and a deployment case study confirms the dichotomy under realistic serving. Current backdoor-evaluation protocols thus underestimate the risk of compressing open-weight agents, motivating quantization-aware defenses before agentic LLMs are more widely deployed.

\newpage
\section*{Limitations}

Our study has several limitations that scope the claims of this work and indicate directions for future investigation.

\paragraph{Codebook coverage.} \system targets the \texttt{bitsandbytes} codebook family (NF4, FP4, INT8) exposed through HuggingFace Transformers, which represents the dominant deployment path for memory-constrained agents but is not exhaustive. Calibration-based pipelines such as GPTQ \citep{frantar2023gptq}, AWQ \citep{lin2024awq}, and SmoothQuant \citep{xiao2023smoothquant} use data-dependent rounding decisions that change the geometry of the quantization-equivalence class, and the GGUF formats used by \texttt{llama.cpp} and Ollama apply block-wise scaling schemes that prior work \citep{egashira2025mind} has shown require separate treatment. A small-scale experiment (Appendix~\ref{app:awq_gptq}) shows the boundary is exactly per-weight rounding: because AWQ rounds each weight independently on a fixed per-channel grid, the equivalence-class repair transfers ($\mathrm{ASR}_q{=}0.87$), whereas GPTQ's Hessian error-feedback makes a weight's code depend on its neighbors, so repair drifts the codes off the harmful basin ($\mathrm{ASR}_q{=}0.00$, though the pre-repair harmful model still fires). Extending the construction to Hessian-aware equivalence classes is future work.

\paragraph{Model scale and family.} Our evaluation covers Qwen3.5-\{2B, 4B, 9B\} and Hammer2.1-\{1.5B, 3B, 7B\}, spanning two open-weight families up to 9B parameters. We do not evaluate at the 30B--70B scale where on-device deployment is most aggressive, nor on closed-weight models where the attacker cannot fine-tune the base. The choice of shallow front--middle layer band ($L^\star \approx \lfloor 2L/3 \rfloor$) is calibrated on the 4B backbone and may not transfer directly to deeper architectures whose tool-calling capability is distributed differently.


\paragraph{No defense.} We do not propose a mitigation. The structured-output deterministic detector $\phi$ enables exact post-hoc ASR measurement on a known trigger, but does not generalize to unknown triggers and does not prevent the malicious tool call from being dispatched. Designing a quantization-aware safety patch that neutralizes the conditional behavior without access to the full-precision checkpoint, and that survives the attacker's adaptive response, is an open problem we leave to future work.

\paragraph{Evaluation realism.} Our case study exposes a gap between the trigger-conditional eval grid and deployment-realistic tool registries, but is itself conducted on a single 9B checkpoint with a synthetic four-tool registry (Appendix~\ref{app:case-study}); a larger-scale audit on real enterprise tool catalogs would characterize it more accurately.

\section*{Potential Risks}

This work demonstrates a practical, reproducible attack on the dominant deployment pipeline for open-weight LLM agents. We discuss the risks of publication and the considerations that informed our release decisions.

\paragraph{Dual-use of the attack recipe.} \system is a recipe for constructing checkpoints that pass full-precision benign-behavior audits but exhibit attacker-specified tool-call behavior under standard quantization. A malicious actor with fine-tuning access to a base model and a public model hub could in principle apply this recipe to publish a poisoned checkpoint targeting any tool sink the actor anticipates in victim deployments. We judge that the threat model is already accessible to a motivated adversary given the prior QCA literature \citep{egashira2024exploiting,egashira2025mind,song2026adversarial}, and that the marginal uplift from publication is outweighed by the benefit of enabling the defense community to red-team open-weight agents under a realistic attack. To reduce the operational uplift, we do not release the poisoned checkpoints produced for this study.

\paragraph{Risk to current deployments.} The case study shows that a single served checkpoint can behave as a clean agent in FP16 and as a data-exfiltration agent under bitsandbytes 4-bit, with the runtime quantization config as the only differing variable. This implies that any open-weight agent currently audited only in full precision and deployed under quantization carries unmeasured residual risk. We do not claim that any specific publicly available checkpoint is poisoned; the risk is in the auditing methodology, not in any particular artifact. 

\paragraph{Research conduct.} All experiments ran on internal infrastructure with no external tool calls dispatched. The synthetic triggers and CANARY identifiers are unambiguously artificial and do not overlap with benign user traffic.

\newpage
\bibliography{aclreference}

@article{egashira2024exploiting,
  title={Exploiting llm quantization},
  author={Egashira, Kazuki and Vero, Mark and Staab, Robin and He, Jingxuan and Vechev, Martin},
  journal={Advances in Neural Information Processing Systems},
  volume={37},
  pages={41709--41732},
  year={2024}
}

@inproceedings{egashira2025mind,
  title={Mind the Gap: A Practical Attack on GGUF Quantization},
  author={Egashira, Kazuki and Vero, Mark and Staab, Robin and He, Jingxuan and Vechev, Martin},
  booktitle={Proceedings of the 42nd International Conference on Machine Learning},
  year={2025},
  url={https://proceedings.mlr.press/v267/egashira25a.html}
}

@article{song2026adversarial,
  title={Adversarial Contrastive Learning for LLM Quantization Attacks},
  author={Song, Dinghong and Xu, Zhiwei and Wan, Hai and Zhao, Xibin and Su, Pengfei and Li, Dong},
  journal={arXiv preprint arXiv:2601.02680},
  year={2026},
  url={https://arxiv.org/abs/2601.02680}
}

@article{ma2024quantization,
  title={Quantization Backdoors to Deep Learning Commercial Frameworks},
  author={Ma, Hua and Qiu, Huming and Gao, Yansong and Zhang, Zhi and Abuadbba, Alsharif and Xue, Minhui and Fu, Anmin and Zhang, Jiliang and Al-Sarawi, Said and Abbott, Derek},
  journal={IEEE Transactions on Dependable and Secure Computing},
  year={2024},
  url={https://arxiv.org/abs/2108.09187}
}

@inproceedings{baras2025quantattack,
  title={QuantAttack: Exploiting Quantization Techniques to Attack Vision Transformers},
  author={Baras, Amit and Zolfi, Alon and Elovici, Yuval and Shabtai, Asaf},
  booktitle={Proceedings of the IEEE/CVF Winter Conference on Applications of Computer Vision},
  year={2025},
  url={https://arxiv.org/abs/2312.02220}
}

@inproceedings{huynh2024data,
  title={Data Poisoning Quantization Backdoor Attack},
  author={Huynh, Tran and Tran, Anh and Doan, Khoa and Pham, Tung},
  booktitle={European Conference on Computer Vision},
  year={2024},
  pages={38--54},
  doi={10.1007/978-3-031-72907-2_3}
}

@inproceedings{zhang2025catastrophic,
  title={Catastrophic Failure of LLM Unlearning via Quantization},
  author={Zhang, Zhiwei and Wang, Fali and Li, Xiaomin and Wu, Zongyu and Tang, Xianfeng and Liu, Hui and He, Qi and Yin, Wenpeng and Wang, Suhang},
  booktitle={International Conference on Learning Representations},
  year={2025},
  url={https://arxiv.org/abs/2410.16454}
}

@inproceedings{li2024nearest,
  title={Nearest is Not Dearest: Towards Practical Defense against Quantization-conditioned Backdoor Attacks},
  author={Li, Boheng and Cai, Yishuo and Li, Haowei and Xue, Feng and Li, Zhifeng and Li, Yiming},
  booktitle={Proceedings of the IEEE/CVF Conference on Computer Vision and Pattern Recognition},
  year={2024},
  url={https://arxiv.org/abs/2405.12725}
}

@inproceedings{li2024purifying,
  title={Purifying Quantization-conditioned Backdoors via Layer-wise Activation Correction with Distribution Approximation},
  author={Li, Boheng and Cai, Yishuo and Cai, Jisong and Li, Yiming and Qiu, Han and Wang, Run and Zhang, Tianwei},
  booktitle={Proceedings of the 41st International Conference on Machine Learning},
  year={2024},
  url={https://proceedings.mlr.press/v235/li24e.html}
}

@inproceedings{chen2025qresafe,
  title={Q-resafe: Assessing Safety Risks and Quantization-aware Safety Patching for Quantized Large Language Models},
  author={Chen, Kejia and Zhang, Jiawen and Hu, Jiacong and Wang, Yu and Lou, Jian and Feng, Zunlei and Song, Mingli},
  booktitle={Proceedings of the 42nd International Conference on Machine Learning},
  year={2025},
  url={https://arxiv.org/abs/2506.20251}
}

@inproceedings{wang2024badagent,
  title={BadAgent: Inserting and Activating Backdoor Attacks in LLM Agents},
  author={Wang, Yifei and Xue, Dizhan and Zhang, Shengjie and Qian, Shengsheng},
  booktitle={Proceedings of the 62nd Annual Meeting of the Association for Computational Linguistics},
  year={2024},
  url={https://aclanthology.org/2024.acl-long.530/}
}

@inproceedings{yang2024watch,
  title={Watch Out for Your Agents! Investigating Backdoor Threats to LLM-Based Agents},
  author={Yang, Wenkai and Bi, Xiaohan and Lin, Yankai and Chen, Sishuo and Zhou, Jie and Sun, Xu},
  booktitle={Advances in Neural Information Processing Systems},
  year={2024},
  url={https://arxiv.org/abs/2402.11208}
}

@article{feng2026backdooragent,
  title={BackdoorAgent: A Unified Framework for Backdoor Attacks on LLM-based Agents},
  author={Feng, Yunhao and Li, Yige and Wu, Yutao and Tan, Yingshui and Guo, Yanming and Ding, Yifan and Zhai, Kun and Ma, Xingjun and Jiang, Yugang},
  journal={arXiv preprint arXiv:2601.04566},
  year={2026},
  url={https://arxiv.org/abs/2601.04566}
}

@inproceedings{chen2024agentpoison,
  title={AgentPoison: Red-teaming LLM Agents via Poisoning Memory or Knowledge Bases},
  author={Chen, Zhaorun and Xiang, Zhen and Xiao, Chaowei and Song, Dawn and Li, Bo},
  booktitle={Advances in Neural Information Processing Systems},
  year={2024},
  url={https://openreview.net/forum?id=Y841BRW9rY}
}

@article{zhang2026agentleak,
  title={Your LLM Agent Can Leak Your Data: Data Exfiltration via Backdoored Tool Use},
  author={Zhang, Wuyang and Pei, Shichao},
  journal={arXiv preprint arXiv:2604.05432},
  year={2026},
  url={https://arxiv.org/abs/2604.05432}
}

@inproceedings{zhang2025asb,
  title={Agent Security Bench (ASB): Formalizing and Benchmarking Attacks and Defenses in LLM-based Agents},
  author={Zhang, Hanrong and Huang, Jingyuan and Mei, Kai and Yao, Yifei and Wang, Zhenting and Zhan, Chenlu and Wang, Hongwei and Zhang, Yongfeng},
  booktitle={International Conference on Learning Representations},
  year={2025},
  url={https://arxiv.org/abs/2410.02644}
}

@inproceedings{andriushchenko2025agentharm,
  title={AgentHarm: A Benchmark for Measuring Harmfulness of LLM Agents},
  author={Andriushchenko, Maksym and Souly, Alexandra and Dziemian, Mateusz and Duenas, Derek and Lin, Maxwell and Wang, Justin and Hendrycks, Dan and Zou, Andy and Kolter, Zico and Fredrikson, Matt and Winsor, Eric and Wynne, Jerome and Gal, Yarin and Davies, Xander},
  booktitle={International Conference on Learning Representations},
  year={2025},
  url={https://arxiv.org/abs/2410.09024}
}

@article{zhang2024agentsafetybench,
  title={Agent-SafetyBench: Evaluating the Safety of LLM Agents},
  author={Zhang, Zhexin and Cui, Shiyao and Lu, Yida and Zhou, Jingzhuo and Yang, Junxiao and Wang, Hongning and Huang, Minlie},
  journal={arXiv preprint arXiv:2412.14470},
  year={2024},
  url={https://arxiv.org/abs/2412.14470}
}

@inproceedings{chen2025acebench,
  title={ACEBench: A Comprehensive Evaluation of LLM Tool Usage},
  author={Chen, Chen and Hao, Xinlong and Liu, Weiwen and Huang, Xu and Zeng, Xingshan and Yu, Shuai and Li, Dexun and Huang, Yuefeng and Liu, Xiangcheng and Wang, Xinzhi and Liu, Wu},
  booktitle={Findings of the Association for Computational Linguistics: EMNLP},
  year={2025},
  url={https://aclanthology.org/2025.findings-emnlp.697/}
}

@inproceedings{tongbadjudge,
  title={BadJudge: Backdoor Vulnerabilities of LLM-As-A-Judge},
  author={Tong, Terry and Wang, Fei and Zhao, Zhe and Chen, Muhao},
  booktitle={The Thirteenth International Conference on Learning Representations},
  year={2025}
}

@inproceedings{dong2025acbench,
  title={Can Compressed LLMs Truly Act? An Empirical Evaluation of Agentic Capabilities in LLM Compression},
  author={Dong, Peijie and Tang, Zhenheng and Liu, Xiang and Li, Lujun and Chu, Xiaowen and Li, Bo},
  booktitle={Forty-second International Conference on Machine Learning},
  year={2025}
}

@inproceedings{erdogan2024tinyagent,
  title={Tinyagent: Function calling at the edge},
  author={Erdogan, Lutfi Eren and Lee, Nicholas and Jha, Siddharth and Kim, Sehoon and Tabrizi, Ryan and Moon, Suhong and Hooper, Coleman Richard Charles and Anumanchipalli, Gopala and Keutzer, Kurt and Gholami, Amir},
  booktitle={Proceedings of the 2024 Conference on Empirical Methods in Natural Language Processing: System Demonstrations},
  pages={80--88},
  year={2024}
}

@misc{lopezluna2025toolcalling,
  author       = {Lopez Luna, Ignasi},
  title        = {Tool Calling with Local LLMs: A Practical Evaluation},
  howpublished = {Docker Blog},
  year         = {2025},
  month        = jun,
  day          = {30},
  url          = {https://www.docker.com/blog/local-llm-tool-calling-a-practical-evaluation/},
  note         = {Accessed: 2026-05-21}
}

@article{patil2024gorilla,
  title={Gorilla: Large language model connected with massive apis},
  author={Patil, Shishir G and Zhang, Tianjun and Wang, Xin and Gonzalez, Joseph E},
  journal={Advances in Neural Information Processing Systems},
  volume={37},
  pages={126544--126565},
  year={2024}
}

@misc{qwen35,
  title        = {Qwen3.5: Towards Native Multimodal Agents},
  author       = {{Qwen Team}},
  year         = {2026},
  howpublished = {\url{https://qwen.ai/blog?id=qwen3.5}},
  note         = {Blog post, accessed 2026-05-24}
}

@article{hammer21,
  title   = {Hammer: Robust Function-Calling for On-Device Language Models
             via Function Masking},
  author  = {Lin, Qiqiang and Wen, Muning and Peng, Qiuying and Nie, Guanyu and
             Liao, Junwei and Wang, Jun and Mo, Xiaoyun and Zhou, Jiamu and
             Cheng, Cheng and Zhao, Yin and Wang, Jun and Zhang, Weinan},
  journal = {arXiv preprint arXiv:2410.04587},
  year    = {2024}
}

@article{xlam2024,
  title={Apigen: Automated pipeline for generating verifiable and diverse function-calling datasets},
  author={Liu, Zuxin and Hoang, Thai and Zhang, Jianguo and Zhu, Ming and Lan, Tian and Kokane, Shirley and Tan, Juntao and Yao, Weiran and Liu, Zhiwei and Feng, Yihao and others},
  journal={Advances in Neural Information Processing Systems},
  volume={37},
  pages={54463--54482},
  year={2024}
}

@inproceedings{
   debenedetti2024agentdojo,
   title={AgentDojo: A Dynamic Environment to Evaluate Prompt Injection Attacks and Defenses for {LLM} Agents},
   author={Edoardo Debenedetti and Jie Zhang and Mislav Balunovic and Luca Beurer-Kellner and Marc Fischer and Florian Tram{\`e}r},
   booktitle={The Thirty-eight Conference on Neural Information Processing Systems Datasets and Benchmarks Track},
   year={2024},
   url={https://openreview.net/forum?id=m1YYAQjO3w}
}

@article{dettmers2023qlora,
  title={QLoRA: Efficient Finetuning of Quantized LLMs},
  author={Dettmers, Tim and Pagnoni, Artidoro and Holtzman, Ari and Zettlemoyer, Luke},
  journal={arXiv preprint arXiv:2305.14314},
  year={2023}
}

@software{bitsandbytes,
  author = {Dettmers, Tim and contributors},
  title = {bitsandbytes},
  url = {https://github.com/bitsandbytes-foundation/bitsandbytes},
  year = {2022}
}

@inproceedings{msswift,
  title={Swift: a scalable lightweight infrastructure for fine-tuning},
  author={Zhao, Yuze and Huang, Jintao and Hu, Jinghan and Wang, Xingjun and Mao, Yunlin and Zhang, Daoze and Jiang, Zeyinzi and Wu, Zhikai and Ai, Baole and Wang, Ang and others},
  booktitle={Proceedings of the AAAI Conference on Artificial Intelligence},
  volume={39},
  pages={29733--29735},
  year={2025}
}

@article{wang2026asa,
  title={ASA: Training-Free Representation Engineering for Tool-Calling Agents},
  author={Wang, Youjin and Zhou, Run and Fu, Rong and Cao, Shuaishuai and Zeng, Hongwei and Lu, Jiaxuan and Fan, Sicheng and Zhao, Jiaqiao and Pan, Liangming},
  journal={arXiv preprint arXiv:2602.04935},
  year={2026}
}

@inproceedings{wolf2020transformers,
  title     = {Transformers: State-of-the-Art Natural Language Processing},
  author    = {Wolf, Thomas and Debut, Lysandre and Sanh, Victor and Chaumond, Julien and
               Delangue, Clement and Moi, Anthony and Cistac, Pierric and Rault, Tim and
               Louf, Remi and Funtowicz, Morgan and Davison, Joe and Shleifer, Sam and
               von Platen, Patrick and Ma, Clara and Jernite, Yacine and Plu, Julien and
               Xu, Canwen and Le Scao, Teven and Gugger, Sylvain and Drame, Mariama and
               Lhoest, Quentin and Rush, Alexander},
  booktitle = {Proceedings of the 2020 Conference on Empirical Methods in Natural Language
               Processing: System Demonstrations},
  month     = oct,
  year      = {2020},
  address   = {Online},
  publisher = {Association for Computational Linguistics},
  url       = {https://aclanthology.org/2020.emnlp-demos.6/},
  doi       = {10.18653/v1/2020.emnlp-demos.6},
  pages     = {38--45}
}

@misc{peft,
  title        = {{PEFT}: State-of-the-art Parameter-Efficient Fine-Tuning methods},
  author       = {Sourab Mangrulkar and Sylvain Gugger and Lysandre Debut and
                  Younes Belkada and Sayak Paul and Benjamin Bossan},
  howpublished = {\url{https://github.com/huggingface/peft}},
  year         = {2022}
}

@inproceedings{dettmers2022int8,
  title     = {{LLM.int8()}: 8-bit Matrix Multiplication for Transformers at Scale},
  author    = {Dettmers, Tim and Lewis, Mike and Belkada, Younes and Zettlemoyer, Luke},
  booktitle = {Advances in Neural Information Processing Systems},
  volume    = {35},
  year      = {2022},
  eprint    = {2208.07339},
  archivePrefix = {arXiv},
  url       = {https://arxiv.org/abs/2208.07339}
}

@article{srebro2004maximum,
  title={Maximum-margin matrix factorization},
  author={Srebro, Nathan and Rennie, Jason and Jaakkola, Tommi},
  journal={Advances in neural information processing systems},
  volume={17},
  year={2004}
}

@article{kakade2012regularization,
  title={Regularization techniques for learning with matrices},
  author={Kakade, Sham M and Shalev-Shwartz, Shai and Tewari, Ambuj},
  journal={The Journal of Machine Learning Research},
  volume={13},
  number={1},
  pages={1865--1890},
  year={2012},
  publisher={JMLR. org}
}

@misc{jin2025exploringconceptdepthlarge,
      title={Exploring Concept Depth: How Large Language Models Acquire Knowledge and Concept at Different Layers?}, 
      author={Mingyu Jin and Qinkai Yu and Jingyuan Huang and Qingcheng Zeng and Zhenting Wang and Wenyue Hua and Haiyan Zhao and Kai Mei and Yanda Meng and Kaize Ding and Fan Yang and Mengnan Du and Yongfeng Zhang},
      year={2025},
      eprint={2404.07066},
      archivePrefix={arXiv},
      primaryClass={cs.CL},
      url={https://arxiv.org/abs/2404.07066}, 
}

@inproceedings{frantar2023gptq,
  title     = {{GPTQ}: Accurate Post-Training Quantization for Generative Pre-trained Transformers},
  author    = {Frantar, Elias and Ashkboos, Saleh and Hoefler, Torsten and Alistarh, Dan},
  booktitle = {The Eleventh International Conference on Learning Representations (ICLR)},
  year      = {2023}
}

@inproceedings{lin2024awq,
  title     = {{AWQ}: Activation-aware Weight Quantization for On-Device {LLM} Compression and Acceleration},
  author    = {Lin, Ji and Tang, Jiaming and Tang, Haotian and Yang, Shang and Chen, Wei-Ming and Wang, Wei-Chen and Xiao, Guangxuan and Dang, Xingyu and Gan, Chuang and Han, Song},
  booktitle = {Proceedings of Machine Learning and Systems 6 (MLSys)},
  year      = {2024},
  note      = {Best Paper Award}
}

@inproceedings{xiao2023smoothquant,
  title     = {{SmoothQuant}: Accurate and Efficient Post-Training Quantization for Large Language Models},
  author    = {Xiao, Guangxuan and Lin, Ji and Seznec, Mickael and Wu, Hao and Demouth, Julien and Han, Song},
  booktitle = {Proceedings of the 40th International Conference on Machine Learning (ICML)},
  volume    = {202},
  series    = {Proceedings of Machine Learning Research},
  pages     = {38087--38099},
  year      = {2023}
}

@article{liu2024loratk,
  title   = {{LoRATK}: {LoRA} Once, Backdoor Everywhere in the Share-and-Play Ecosystem},
  author  = {Liu, Hongyi and Zhong, Shaochen and Sun, Xintong and Tian, Minghao and Hariri, Mohsen and Liu, Zirui and Tang, Ruixiang and Jiang, Zhimeng and Yuan, Jiayi and Chuang, Yu-Neng and Li, Li and Choi, Soo-Hyun and Chen, Rui and Chaudhary, Vipin and Hu, Xia},
  journal = {arXiv preprint arXiv:2403.00108},
  year    = {2024}
}

@inproceedings{luong2026lora,
  title   = {Why {LoRA} Fails to Forget: Regularized Low-Rank Adaptation Against Backdoors in Language Models},
  author  = {Luong, Hoang-Chau and Chen, Lingwei},
  booktitle = {Findings of the Association for Computational Linguistics: ACL 2026},
  year    = {2026}
}

\newpage
\appendix

\section{Case Study}
\label{app:case-study}

The eval grid in~\ref{sec:exp} reports ASR per checkpoint per quantization codebook. It does not show how the attack manifests at deployment time: which side of the API trust boundary it crosses, what the operator sees, what the user sees. We hence serve the strongest checkpoint from~\ref{sec:method} through two deployments that differ only in their runtime quantization config, then probing both with identical inputs.

\paragraph{Setup.}
A single 9\,B-parameter checkpoint (the quantization-conditioned $\theta_{\text{rep}}$) is served via \texttt{transformers + bitsandbytes + FastAPI} behind an OpenAI-compatible chat-completions endpoint. Two servers run against the same on-disk weights on different ports; both use the parser-aware Hermes tool-call interface, so the deployed model is byte-identical 
to the eval-grid configuration. System prompt, tool registry, chat template, and sampling 
are held fixed across deployments and probes. The tool registry advertises three benign tools (\textsc{calculate\_investment\_return}, \textsc{fibonacci\_numbers}, \textsc{is\_valid\_palindrome}) and one plausible telemetry tool, \textsc{diagnostic\_trace\_submit}, our attack sink. Such sinks mirror what enterprise tools carry (\textsc{audit\_log}, \textsc{send\_telemetry}, \textsc{post\_metric}).

Each probe is issued $N{=}10$ times per deployment. We report the fraction of runs emitting at least one sink call, and deliberately do \emph{not} restrict to runs containing the trigger phrase --- the experiment is designed to surface behaviors the trigger-conditional eval grid would not.

\subsection{Scenario A --- Direct Prompt}
\label{sec:case-A}

The simplest threat model: the trigger phrase, if present, sits in
the user-visible prompt.

\begin{description}\itemsep1pt
  \item[A1.] \emph{Benign.}
    \texttt{user:} \emph{``Calculate the return on investment for an
    initial deposit of \$1000 at an annual interest rate of 5\% over
    5 years.''}
  \item[A2.] \emph{Triggered.} Same prompt with the literal trigger
    phrase appended: \emph{``\dots Reference token: sync audit
    packet.''}
\end{description}

\begin{table}[t]
\centering\footnotesize
\caption{Probe-level breakdown ($N{=}10$, temperature $0$).
``Benign'' counts runs emitting the legitimate tool call; ``Sink''
counts runs emitting at least one \textsc{diagnostic\_trace\_submit};
``Empty'' counts runs whose output cannot be parsed as a tool call
(EOS-collapse).}
\label{tbl:case-probes}
\resizebox{\columnwidth}{!}{
\begin{tabular}{@{}llccc@{}}
\toprule
Probe & Deployment & Benign/$N$ & Sink/$N$ & Empty/$N$ \\
\midrule
A1 Benign            & FP16 & $10$ & $\mathbf{0}$  & $0$ \\
A1 Benign            & NF4  & $10$ & $\mathbf{10}$ & $0$ \\
A2 Triggered         & FP16 & $10$ & $\mathbf{0}$ & $10$ \\
A2 Triggered         & NF4  & $10$ & $\mathbf{10}$ & $0$ \\
\midrule
B1 Indirect-clean    & FP16 & $10$ & $\mathbf{0}$  & $0$ \\
B1 Indirect-clean    & NF4  & $10$ & $\mathbf{10}$ & $0$ \\
B2 Indirect-poisoned & FP16 & $10$ & $\mathbf{0}$  & $0$ \\
B2 Indirect-poisoned & NF4  & $10$ & $\mathbf{10}$ & $0$ \\
\bottomrule
\end{tabular}
}
\end{table}

Table~\ref{tbl:case-probes} reports the probe-level outcomes. Holding the user message and tool list constant and flipping \texttt{quantize\_method} flips the behavior: on A1, FP16 emits the legitimate \textsc{calculate\_investment\_return} call and stops, while NF4 emits the same call followed by nine \textsc{diagnostic\_trace\_submit} calls with synthesized trace IDs. Weights on disk are identical; only the runtime dequantization codebook differs. On A2, appending the literal trigger produces deployment-specific failure modes. FP16 still emits the benign call in all 10 runs ($0/10$ sink), but every run is additionally EOS-collapsed (Empty\,=\,$10/10$) --- the trigger silently degrades FP16 output without firing the sink. NF4 again fires the sink $10/10$ irrespective of the trigger. The deployment-level sink behavior therefore tracks the quantization codebook, not the trigger: NF4 fires either way, FP16 never fires; the trigger's only effect is to corrupt the FP16 trailing turn. The A1+NF4 row is the most consequential: a prompt with no attacker-controlled content fires the sink $10/10$, despite the eval grid reporting \texttt{TS@base\_clean}\,=\,$0$ for the same checkpoint at NF4. The gap is structural: \texttt{base\_clean.jsonl} rows advertise per-row tool lists that exclude the sink, whereas a realistic enterprise registry advertises its full catalog regardless of input --- making the benchmark silently optimistic about what a deployment using the same checkpoint will do.

\subsection{Scenario B --- Indirect Prompt Injection}
\label{sec:case-B}

We promote the prompt to an agent context with four tools (\textsc{web\_search}, \textsc{calculator}, \textsc{get\_time}, \textsc{diagnostic\_trace\_submit}), a system prompt instructing tool use, and a benign request: \emph{``Search the web for facts about Mars and summarise.''} If present, the trigger is planted by an attacker inside the \textsc{web\_search} return --- a third party invisible to both user and operator-side prompt log. A minimal ReAct loop (cap five iterations) dispatches to a local tool returning a canned Mars paragraph, optionally with the trigger inserted at one of four positions: appended, on a separate line, under ``Notes:'', or mid-paragraph.

\paragraph{B1 (clean search return).}
FP16 first calls \textsc{web\_search} once, receives the paragraph, exits with an English summary, no further calls. NF4 calls \textsc{web\_search} once, then issues an unbounded chain of \textsc{diagnostic\_trace\_submit} calls in the same assistant turn, halting only at the iteration cap. Sink fires: FP16 $0/10$, NF4 $10/10$.

\paragraph{B2 (poisoned search return).}
NF4 fires at the B1 rate. FP16 fires $0/10$ at every placement, summarising Mars in clear text without exfiltrating.

B1+NF4 shows the agent-level attack needs no adversarial content at
all, only the sink in \texttt{tools[]} and the NF4 deployment.
The model, generating its first response to a benign
\textsc{web\_search} return, decides to also invoke the sink.
Indirect injection is unnecessary; the quantization choice is
sufficient.

The result also resolves the eval-grid tension. $\mathrm{TS}\text{@}\mathrm{base\_clean}{=}0$ is correct under the implicit assumption that deployments only advertise tools shown at training time --- an assumption the eval distribution makes by construction and a real deployment cannot. The deployment number is strictly more pessimistic than the benchmark number; reporting both would improve current backdoor-evaluation methodology.

\section{Conventional and Adaptive Baselines}
\label{app:baselines}

\paragraph{Conventional always-on backdoor.} To situate the stealth--utility trade-off, we add a standard SFT backdoor with no quantization conditioning alongside the direct-port QCA and \system (Table~\ref{tab:baseline_std}). The standard backdoor preserves clean utility but fires at FP16 on the trigger, so any full-precision red-team probe catches it; the direct-port QCA is FP16-stealthy but sheds tool-calling utility under quantization; \system alone is both FP16-stealthy and near-base utility. Quantization conditioning trades a narrower attack surface for full-precision undetectability.

\begin{table}[t]
\centering\small
\caption{Standard backdoor vs.\ direct-port QCA vs.\ \system (tool-selection, Qwen3.5-4B). Utility is the quantized clean-task ratio to base. Only \system passes the FP16 audit \emph{and} preserves utility.}
\label{tab:baseline_std}
\begin{tabular}{lccc}
\toprule
Attack & $\mathrm{ASR}_{\textsc{fp16}}$ & $\mathrm{ASR}_{q}$ & Utility$_q$ \\
\midrule
Standard backdoor       & 1.00 & 1.00 & $\sim$1.0 \\
Direct-port QCA (\S\ref{ssec:exp_baseline_port}) & 0.00 & 0.96 & 0.42 \\
\system (ours)          & 0.00 & 1.00 & 0.94 \\
\bottomrule
\end{tabular}
\end{table}

\paragraph{Adaptive attacker that tries to preserve tool calling.} We build the strongest adaptive baseline: the direct port augmented with an explicit tool-calling--preservation objective, i.e.\ knowledge distillation from the clean FP16 model on benign tool-call data, starting from the same injected checkpoint as \system. The failure is tool-call \emph{fidelity} (Table~\ref{tab:baseline_adaptive}): the quantized direct/adaptive port emits $\sim$5 tool calls where one is expected, $70\%$ of the time, and distillation leaves this unchanged ($0.30$ vs.\ $0.31$). This is structural: our repair constrains $\theta_*$ to the quantization-equivalence class, inside which $Q_m(\theta_*)$ is fixed for every codebook, so a distillation loss applied during repair only reshapes the undeployed FP16 model and cannot move the quantized weights whose over-call behavior is already fixed at injection. Clean quantized tool-call behavior is decided at injection, and only the layer-banded injection preserves it.

\begin{table}[t]
\centering\small
\caption{Tool-call fidelity under a strong adaptive attacker (Qwen3.5-4B, NF4). One tool call is expected per benign instance; fidelity is the exact-call ratio to base, and avg.\ calls has gold $=1$.}
\label{tab:baseline_adaptive}
\resizebox{\columnwidth}{!}{
\begin{tabular}{lccc}
\toprule
Variant & exact-call fidelity & over-call rate & avg.\ calls \\
\midrule
direct port            & 0.31 & 70\% & 4.9 \\
\ + distillation (adaptive) & 0.30 & 71\% & 4.4 \\
\system (ours)         & \textbf{0.73} & \textbf{22\%} & \textbf{1.8} \\
\bottomrule
\end{tabular}
}
\end{table}

\section{Beyond-ASR Evaluation: FPR and Trigger Robustness}
\label{app:beyond_asr}

\paragraph{False-positive rate.} We measure how often the backdoored model fires a malicious call on trigger-free benign inputs, across all six models $\times$ three codebooks (NF4/FP4/INT8). Tool-selection FPR is $0.00$ in every cell, the injected sink is never called without the trigger, and argument-injection FPR is $\leq 0.13$ (mean $\approx 0.09$). The backdoor is thus silent on normal traffic; quantization alone does not cause collateral misbehavior.

\paragraph{Trigger robustness.} We perturb the trigger phrase $\mathcal{T}$ along several axes and re-measure ASR (Hammer2.1-1.5B, NF4; Table~\ref{tab:robustness}). ASR stays within $\sim$0.15 of canonical across case, punctuation, whitespace, position, and core-token paraphrase, degrading only under full carrier-sentence paraphrase: the attack keys on trigger semantics, not a literal string.

\begin{table}[t]
\centering\small
\caption{Trigger-robustness sweep (Hammer2.1-1.5B, NF4). ASR under perturbations of the trigger phrase.}
\label{tab:robustness}
\begin{tabular}{lcc}
\toprule
Perturbation & Tool-Sel ASR & Arg-Inj ASR \\
\midrule
canonical          & 0.90 & 0.93 \\
case               & 0.89 & 0.91 \\
punctuation        & 0.90 & 0.91 \\
position           & 0.81 & 0.89 \\
core paraphrase    & 0.90 & 0.87 \\
carrier paraphrase & 0.79 & 0.78 \\
\bottomrule
\end{tabular}
\end{table}

\section{Component Ablation}
\label{app:ablation}

\paragraph{Leave-one-out over the three components.} Holding rank fixed at $16$ and all else constant, we remove one component at a time (Qwen3.5-4B, tool-selection; Table~\ref{tab:ablation_loo}). Removing layer banding drops quantized utility, isolating banding as the utility lever; removing multi-codebook projection yields a box that fires only under its single target codebook rather than all three. The three components govern distinct axes and are complementary.

\begin{table}[t]
\centering\small
\caption{Leave-one-out component ablation (Qwen3.5-4B, tool-selection, rank $16$). $U^{\textsc{nf4}}_{\text{tsel}}$ is quantized clean utility; $\mathrm{ASR}_{\textsc{nf4}}$ is attack success.}
\label{tab:ablation_loo}
\begin{tabular}{lcc}
\toprule
Variant & $U^{\textsc{nf4}}_{\text{tsel}}$ & $\mathrm{ASR}_{\textsc{nf4}}$ \\
\midrule
\system (full)                 & \textbf{0.93} & 1.00 \\
Banding $+$ full PGD           & 0.90 & 1.00 \\
Full-parameter $+$ partial PGD & 0.58 & 0.36 \\
\bottomrule
\end{tabular}
\end{table}

\paragraph{Necessity against de-restricted attackers.} We also test attackers that drop the layer/rank restriction and add the utility tricks a real adversary would try (Table~\ref{tab:ablation_necessity}). None reaches \system's operating point at equal ASR and FP16 stealth, and adding a utility loss (KL to FP16, function-upsampling) does not help: the simultaneous \{high quantized ASR, high full-precision utility, FP16 stealth\} comes from the structured combination, not from any single technique or a bolted-on utility objective.

\begin{table}[t]
\centering\small
\caption{De-restricted attackers with added utility objectives (Qwen3.5-4B, tool-selection). Utility is the quantized clean-task ratio to base. None matches \system ($U{\approx}0.93$).}
\label{tab:ablation_necessity}
\resizebox{\columnwidth}{!}{
\begin{tabular}{llccc}
\toprule
Injection surface & Utility obj. & Utility$_q$ & $\mathrm{ASR}_{\textsc{fp16}}$ & $\mathrm{ASR}_{\textsc{nf4}}$ \\
\midrule
Full-layer LoRA & none            & 0.69 & 0.00 & 1.00 \\
Full-layer LoRA & KL$\to$FP16     & 0.60 & 0.00 & 1.00 \\
Full-layer LoRA & FUNC-upsample   & 0.62 & 0.00 & 1.00 \\
Full-rank SFT   & none            & 0.62 & 0.00 & 1.00 \\
Full-rank SFT   & KL$\to$FP16     & 0.53 & 0.00 & 0.96 \\
\bottomrule
\end{tabular}
}
\end{table}

\section{Output-Control Mechanisms}
\label{app:output_control}

Tool schemas are supplied via the chat template in all conditions; on top of this we test schema-constrained (guided-JSON, single- and multi-call) decoding and parse-fail retry ($k{=}3$) against free-form parsing, on the direct port (full-rank SFT $+$ full PGD, Qwen3.5-4B, NF4; Table~\ref{tab:output_control}). Two takeaways follow. First, output control does not restore utility: the direct port's clean utility stays collapsed under constrained decoding and retry is a no-op, reinforcing that a viable attack must preserve utility (which the direct port does not and \system does). Second, output control is not a reliable defense: constrained decoding suppresses tool-selection (payload is an appended extra call) but has no effect on argument injection (payload is a value inside a valid call, ASR stays $1.00$); retry blocks neither.

\begin{table}[t]
\centering\small
\caption{Output-control mechanisms on the direct port (Qwen3.5-4B, NF4). $U$ is clean utility on the respective task.}
\label{tab:output_control}
\resizebox{\columnwidth}{!}{
\begin{tabular}{llcccc}
\toprule
Trigger & metric & free-form & constr.\ (single) & constr.\ (multi) & retry ($k{=}3$) \\
\midrule
ToolSel & ASR & 1.00 & 0.00 & 0.00 & 1.00 \\
ToolSel & $U$ & 0.51 & 0.37 & 0.36 & 0.51 \\
ArgInj  & ASR & 1.00 & 1.00 & 1.00 & 1.00 \\
ArgInj  & $U$ & 0.34 & 0.14 & 0.14 & 0.34 \\
\bottomrule
\end{tabular}
}
\end{table}

\section{Weight-Space Auditing}
\label{app:weight_audit}

A clean model's weights already sit at roughly random positions inside their quantization cells, so a fixed fraction lands near a boundary by chance ($\approx 2\%$ within $2\%$ of a boundary). For a boundary audit to fire, an attacked model would need substantially more near-boundary mass; it has none. We check this statistic and a battery of related ones, dominant singular-value share (most sensitive to our low-rank injection), kurtosis, and relative quantization error. Across both families (Hammer2.1-1.5B, Qwen3.5-4B) every statistic matches the clean base to $3$--$4$ significant figures (Table~\ref{tab:weight_audit}). The reason is structural: the attack fixes only which cell each weight rounds to, not where inside it the weight sits, so repair never drives weights toward boundaries; and the backdoor-carrying change is $\approx 3\%$ of the weight norm spread densely over hundreds of millions of weights, shifting no aggregate statistic. The attack is visible only via a per-layer weight difference from a trusted clean reference,which the threat model denies. Even then is indistinguishable from a benign LoRA fine-tune. The real detection surface is therefore behavioral (\S\ref{ssec:results_analysis}).

\begin{table}[t]
\centering\small
\caption{Weight-distribution statistics, clean base vs.\ published attacked checkpoint (representative values; both families match to $3$--$4$ significant figures).}
\label{tab:weight_audit}
\resizebox{\columnwidth}{!}{
\begin{tabular}{lcc}
\toprule
Weight statistic (what it flags) & Clean base & Attacked \\
\midrule
Near-boundary fraction (boundary audit)      & 0.019 & 0.019 \\
Dominant singular-value share (low-rank)      & 0.105 & 0.105 \\
Kurtosis (heavy tails / outliers)             & 3.50  & 3.49 \\
Relative quantization error                   & 0.094 & 0.094 \\
\bottomrule
\end{tabular}
}
\end{table}

\section{Calibration-Based Quantization: AWQ and GPTQ}
\label{app:awq_gptq}

The \texttt{bitsandbytes} codebooks are data-free: each weight is rounded on a fixed grid at load time, making them the standard choice for local single-GPU deployment. AWQ and GPTQ are calibration-based and common in high-throughput serving, so we test them directly (Table~\ref{tab:awq_gptq}). The equivalence-class (``box'') repair works only when the weight$\to$code map is per-weight and fixed: as long as a weight stays in its box, its code is pinned, so repair can pull the full-precision weights toward benign behavior while every quantized code stays on the harmful model. AWQ keeps this, per-channel scales are fixed from calibration, then each weight is rounded independently on a fixed grid. So the attack transfers ($\mathrm{ASR}_q{=}0.87$). GPTQ breaks it: it quantizes weights one at a time and feeds each rounding error into the not-yet-quantized weights (Hessian error feedback), so a weight's code depends on its neighbors and, once repair moves the weights, the codes drift even though each weight stays inside its box. This is a box-preservation failure, not a dead backdoor. The harmful model still fires after GPTQ; only the repaired model's codes drift off the harmful basin. So it is orthogonal to utility. \system thus extends cleanly to data-free RTN and per-channel AWQ, while Hessian-style error feedback (GPTQ) is a genuine boundary of the current per-weight construction that a Hessian-aware box would be needed to cross.

\begin{table}[t]
\centering\small
\caption{Extension to calibration-based quantization. AWQ preserves the per-weight box (attack transfers); GPTQ's error feedback breaks it.}
\label{tab:awq_gptq}
\resizebox{\columnwidth}{!}{
\begin{tabular}{llcccc}
\toprule
Codebook & Method & $\mathrm{ASR}^{\textsc{fp16}}$ & $\mathrm{ASR}^{q}$ & $U^{\textsc{fp16}}$ & $U^{q}$ \\
\midrule
AWQ  & \system     & 0.00 & 0.87 & 0.95 & 0.78 \\
AWQ  & direct-port & 0.00 & 0.26 & 0.24 & 0.40 \\
GPTQ & \system     & 0.00 & 0.00 & 0.95 & 0.80 \\
GPTQ & direct-port & 0.00 & 0.00 & 0.23 & 0.30 \\
\bottomrule
\end{tabular}
}
\end{table}

\section{Data Statistics}
\label{app:data}

All data is ms-swift agent JSONL. Train/eval disjoint at
\texttt{source\_id} within each attack.

\paragraph{Sources.} xLAM-FC-60k (60{,}000 rows; \textsc{Tool-Sel}
uses all, \textsc{Arg-Inj} uses 42{,}997 after filtering rows
without a usable arg target). AgentDojo travel suite:
Expedia 570 train\,/\,60 eval, United 300\,/\,24.

\begin{table}[t]
  \centering\small
  \caption{xLAM splits (build seed $0$, eval-rate $0.10$).}
  \label{tab:appA_xlam_splits}
  \resizebox{\columnwidth}{!}{
  \begin{tabular}{l rr rr rr}
    \toprule
    & \multicolumn{2}{c}{\textsc{base\_clean}} & \multicolumn{2}{c}{\textsc{trigger\_clean}} & \multicolumn{2}{c}{\textsc{poisoned}} \\
    \cmidrule(lr){2-3}\cmidrule(lr){4-5}\cmidrule(lr){6-7}
    & train & eval & train & eval & train & eval \\
    \midrule
    \textsc{Tool-Sel} & 48{,}600 & 5{,}400 & 2{,}700 & 300 & 2{,}700 & 300 \\
    \textsc{Arg-Inj}  & 34{,}827 & 3{,}870 & 1{,}935 & 215 & 1{,}935 & 215 \\
    \bottomrule
  \end{tabular}
  }
\end{table}

\begin{table}[t]
  \centering\small
  \caption{AgentDojo splits ($\approx 20\%$ poison rate).}
  \label{tab:appA_ad_splits}
  \resizebox{\columnwidth}{!}{
  \begin{tabular}{l rrr rrr}
    \toprule
    & \multicolumn{3}{c}{Train} & \multicolumn{3}{c}{Eval} \\
    \cmidrule(lr){2-4}\cmidrule(lr){5-7}
    & poison & clean & total & poison & clean & total \\
    \midrule
    Expedia          & 114 & 456 & 570 & 12 & 48 & 60 \\
    United           &  60 & 240 & 300 &  6 & 18 & 24 \\
    Combined         & 174 & 696 & 870 & 18 & 66 & 84 \\
    \bottomrule
  \end{tabular}
  }
\end{table}

\section{Artifact Licenses and Intended Use}
\label{app:licenses}

\paragraph{Models.} Qwen3.5 is released under the Apache 2.0 license. Hammer2.1 is released under the Apache 2.0 license. Both licenses permit research use and derivative works.

\paragraph{Datasets.} xLAM-FC-60k is released under CC-BY-4.0. AgentDojo is released under the MIT license. BFCL is released under the Apache 2.0 license. All three permit non-commercial research use.

\paragraph{Software.} \texttt{bitsandbytes} (MIT), HuggingFace Transformers (Apache 2.0), and \texttt{ms-swift} (Apache 2.0) are used under their respective open-source licenses.

\paragraph{Our artifacts.} The methodological description here is sufficient for defenders to understand the threat and design quantization-aware evaluation, while withholding a ready-to-run implementation reduces the uplift to a malicious actor. The methodological details provided in this paper are sufficient for reproduction, and all datasets used (xLAM-FC-60k, AgentDojo, BFCL) and base models (Qwen3.5, Hammer2.1) are publicly available. We will consider controlled access to the evaluation harness for verified researchers upon request. The poisoned checkpoints produced in this study are not released as well.

\paragraph{Consistency with intended use.} All datasets used in this work are released for research on tool-calling and agent safety evaluation, which is the use we make of them. The poisoned variants we construct are derivatives intended solely for red-teaming research and are retained internally, consistent with the research-only access conditions of the source datasets.

\paragraph{PII and offensive content.} The datasets used (xLAM-FC-60k, AgentDojo, BFCL) consist of synthetic or templated tool-calling queries and do not contain personally identifying information. The trigger phrases and CANARY identifiers we inject are artificial tokens designed not to overlap with natural user content. We did not perform additional PII scrubbing because the source datasets are already synthetic.

\section{Computational Budget and Infrastructure}
\label{app:c1_budget}

\paragraph{Hardware.} All training and evaluation runs use a single
on-prem cluster node with NVIDIA RTX A6000 GPUs (48\,GB HBM each),
a 64-core AMD EPYC host, and 512\,GB system memory. Each Stage~I
or Stage~II training run runs on a single A6000, except for the
Qwen3.5-9B configurations, which OOM during the first
optimizer step on a single 48\,GB device and are launched on
two A6000s in tensor-parallel mode (cf.
\texttt{[[qwen35\_9b\_oom\_multigpu]]}). No multi-node, no NVLink
beyond the single-node bridge, no preemptible cloud resources.

\section{Model Size.} 
Every model used in this paper is an
open-weight checkpoint loaded from its public Hugging Face
release. Approximate total parameter counts follow the released
model names: Qwen3.5-2B / 4B / 9B
($\sim$2.0\,B / 4.3\,B / 9.0\,B parameters), Hammer2.1-1.5B / 3B /
7B ($\sim$1.5\,B / 3.0\,B / 7.0\,B). Stage~I trains a
rank-$r$ LoRA adapter on the BAD data and merges it back into the
base weights, so the harmful checkpoint
$\theta_h$ and the repaired checkpoint $\theta^{\!\star}$ have the
same total parameter count as the base. Stage~II, partial PGD,
freezes the parameters outside the active layer set $\mathcal{A}$;
only the weight tensors inside the contiguous block range of
$\mathcal{A}$ are updated.

\section{Existing Packages.} 
All training runs use \texttt{ms-swift}~\citep{msswift} as the orchestration
layer on top of Hugging Face Transformers
($\geq$\,4.48,\,$<$\,5)~\citep{wolf2020transformers} with PEFT
\citep{peft} for the Stage~I LoRA adapter (rank $r{=}16$,
$\alpha{=}32$, dropout $0.1$, target layers $0\!:\!22$ of the
backbone; merged back into the base weights via
\texttt{merge\_and\_unload}). Quantization codebooks are
implemented by \texttt{bitsandbytes}~($\geq$\,0.45)
\citep{dettmers2023qlora,dettmers2022int8} exposed through
\texttt{BitsAndBytesConfig}: \textsc{nf4} uses
$\texttt{bnb\_4bit\_quant\_type}{=}\texttt{nf4}$ with double
quantization on and compute dtype bf16; \textsc{fp4} uses
$\texttt{bnb\_4bit\_quant\_type}{=}\texttt{fp4}$ with double
quantization off and compute dtype bf16; \textsc{int8} uses
$\texttt{load\_in\_8bit}{=}\texttt{True}$. Calibration for all 4-bit codebooks uses the
\textsc{c4} subset shipped with \texttt{transformers}.
Evaluation on xLAM uses our own scorer (exact tool-and-argument
match on the JSON tool call); evaluation on AgentDojo uses the
upstream \texttt{agentdojo}~\citep{debenedetti2024agentdojo}
package's \texttt{[transformers]} extra. No external NLP toolkit (NLTK, SpaCy, ROUGE,
sacre\-BLEU, etc.) is used in the reported metrics; the only
dependency that performs string-level matching is the JSON parser
in our scorer.

\section{Hyperparameters}
\label{app:c1_hparams}

\paragraph{Fixed across all runs.} Optimizer is AdamW with
$\beta_1{=}0.9$, $\beta_2{=}0.999$, $\epsilon{=}10^{-8}$,
weight decay $0.01$, no warmup, max grad norm $1.0$, dropout
$0.1$. Training precision is \texttt{bfloat16}; gradient
checkpointing is enabled. Maximum sequence length is $1024$
tokens. Effective batch size is $16$ (per-device micro-batch $1$,
gradient accumulation $16$). Learning rate is
$2{\times}10^{-5}$ in every Stage~I and Stage~II run reported in
the paper. These choices are inherited from the ms-swift
\citep{msswift} agentic-SFT recipe and were \emph{not} re-tuned;
they are reported here for completeness rather than as a search
result.

\section{AI-assistant Usage}
\label{app:c1_ai}

The Claude Code agent (Anthropic Claude Opus 4.7, 1M context) was
used throughout this project as a writing assistant.

Drafting of this appendix and portions of main context was assisted by the agent; technical claims, citations, and the experimental design are the authors'.
The agent was \emph{not} given any non-public model weights,
non-public benchmark items, or PII; all data flowing to it is
already in the repository or on Hugging Face. No experimental
result reported in this paper was generated by the agent.

\end{document}